\documentclass[useAMS,usenatbib,referee]{biom}

\usepackage[figuresright]{rotating}
\usepackage{graphicx}
\usepackage{amssymb}
\usepackage{amsmath}
\usepackage{mathtools}
\usepackage{changepage}
\usepackage{float}
\usepackage{bm}
\usepackage{hyperref}

\newcommand{\AR}{\mbox{\tiny{AR}}}

\newcommand{\EN}{\mbox{\tiny{EN}}}

\newcommand\numberthis{\addtocounter{equation}{1}\tag{\theequation}}

\title[Estimating efficacy of measles supplementary immunization activities]{Estimating efficacy of measles supplementary immunization activities via discrete-time modeling of disease incidence time series}

\author
{Tracy Qi Dong\emailx{qd8@uw.edu} \\
Department of Biostatistics, University of Washington, Seattle, WA 98195, USA 
\and
Jon Wakefield\emailx{jonno@uw.edu} \\
Departments of Statistics and Biostatistics, University of Washington, Seattle, WA 98195, USA}

\begin{document}

\pagerange{\pageref{firstpage}--\pageref{lastpage}} 
\pubyear{2020}
\volume{1}
\artmonth{October}

\label{firstpage}

\begin{abstract}

Measles is a significant source of global disease burden and child mortality. Measles vaccination through routine immunization (RI) programs in high-burden settings remains a challenge due to poor health care infrastructure and access. Supplementary immunization activities (SIA) in the form of vaccination campaigns are therefore implemented to prevent measles outbreaks by reducing the size of the susceptible population. The SIA efficacy, defined as the fraction of susceptible population immunized by an SIA, is a critical metric for assessing campaign effectiveness. We propose a discrete-time hidden Markov model for estimating SIA efficacy and forecasting future incidence trends using reported measles incidence data. Our approach extends the time-series susceptible-infected-recovered (TSIR) framework by adding a model component to capture the impact of SIAs on the susceptible population. It also accounts for under-reporting and its associated uncertainty via a two-stage estimation procedure with uncertainty propagation. The proposed model can be used to estimate the underlying susceptible population dynamics, assess how many susceptible people were immunized by past SIAs, and forecast incidence trends in the future under various hypothetical SIA scenarios. We examine model performance via simulations under various levels of under-reporting, and apply the model to analyze monthly reported measles incidence in Benin from 2012 to 2018. 

\end{abstract}

%
%


\maketitle


\section{Introduction} \label{Chap4SecIntro}

Measles is a vaccine-preventable, highly contagious respiratory infection that affects mostly young children \citep{who_measles}. Accelerated immunization activities in recent decades have had a major impact on reducing global measles deaths \citep{patel2019progress}, but disease burden is still high in many low- and middle-income countries (LMICs) where vaccination remains a challenge due to poor health care infrastructure and access \citep{who_measles1}. In high-burden settings, a key strategy for increasing MCV coverage is to conduct supplementary immunization activities (SIAs) in the form of vaccination campaigns, in addition to delivering scheduled vaccination through routine immunization (RI) programs \citep{who_measles1, mri}. During these campaigns, health workers run fixed-post vaccination sites and provide MCV to all children in a pre-specified target age group, regardless of whether they have been vaccinated previously. The aim is to immunize children missed by RI and reduce the susceptible population. 

A common metric used for evaluating SIA effectiveness is the \textit{administrative campaign coverage}. It is calculated as the ratio of the number of MCV doses administered during a campaign to the size of the target population of the campaign. While SIAs usually have high reported campaign coverage, it is unclear how many people are effectively removed from the susceptible population \citep{mbabazi2009achieving}. Another popular indicator of SIA effectiveness is the \textit{SIA coverage among MCV zero-dose children}, which is defined as the proportion of children in a specific age group with no history of receipt of MCV before the SIA who received a dose of MCV during the SIA. \citet{portnoy2018impact} analyzed Demographic and Health Survey (DHS) data from a few low- and middle-income countries (LMICs) to investigate how many children who have not previously received a measles vaccine dose are reached by SIAs. However, their methods require surveys being conducted within 2 years of implementation of SIAs and having a measles SIA indicator variable, which resulted in only 14 out of 111 countries having SIAs and DHS in the years of 2000 onward, to be eligible for the analysis. More recently, \citet{utazi2020geospatial} analyzed the Nigeria 2017--18 post-campaign coverage survey (PCCS) data to estimate the \textit{SIA coverage among MCV zero-dose children} of the campaign. However, high quality PCCS are rare in most LMICs, and the resultant estimates cannot be easily applied to plan future campaigns. 

An attractive measure of SIA effectiveness is the \textit{SIA efficacy}, defined as the fraction of the susceptible population effectively removed via vaccination after a measles SIA. However, estimation of \textit{SIA efficacy} has been a programmatic challenge due to difficulties in estimating the underlying susceptible population. \citet{thakkar2019decreasing} proposed a linear regression approach that incorporates \textit{SIA efficacy} to analyze measles incidence data in Pakistan to optimize campaign timing. While their method demonstrates effective use of \textit{SIA efficacy} in campaign planning, it does not properly account for the uncertainties due to the under-reporting of disease incidence. Specifically, their approach assumes a constant reporting ratio at every time point and ignores any binomial variation in the observed incidence data. The underlying true incidence time series are calculated by simply dividing the observed incidence by a single point estimate of the reporting ratio. As such, the method tends to under-estimate the uncertainties associated with the model parameters and predicted incidence.

In this paper, we extend the time-series susceptible-infected-recovered (TSIR) framework \citep{finkenstadt2000time} and propose a discrete-time hidden Markov model for estimating SIA efficacy using a hybrid of ordinary least squares (OLS) regression with robust standard errors \citep{huber1967behavior, white1980heteroskedasticity} and Markov chain Monte Carlo (MCMC) procedures. The TSIR model was first introduced by \citet{finkenstadt2000time} to provide an approximation for measles time series in large communities where the disease is endemic. It was then extended in a series of subsequent papers \citep{bjornstad2002dynamics, grenfell2002dynamics, glass2003interpreting, morton2005discrete} and has been used to understand measles and rubella transmission in a variety of settings \citep{ferrari2008dynamics, metcalf2011epidemiology, mahmud2017comparative, metcalf2013implications}. A recent review by \citet{wakefield2019spatio} details the theoretical motivation of the framework and compares it with other popular epidemic modelling approaches such as the Epidemic/Endemic (EE) models introduced by \citet{held2005statistical}. More details regarding the TSIR framework are provided in Web Appendix A in the Supporting Information. Our approach extends the existing framework by adding a model component to capture the impact of SIAs on the susceptible population. It also accounts for under-reporting and its associated uncertainty via a two-stage estimation procedure with uncertainty propagation. In addition, our model allows for seasonality in measles transmission and accommodates monthly reported incidence data that are publicly available from the WHO Measles and Rubella Surveillance Database for most member countries \citep{who_measles3}. The proposed model can be used to estimate the underlying susceptible population dynamics, assess how many susceptible people were immunized by past SIAs, and forecast incidence trends in the future under various hypothetical SIA scenarios. 

This paper is structured as follows. In Section \ref{Chap4SecMethod}, we describe the model and the inference details. A simulation study in Section \ref{Chap4SecSim} considers various levels of under-reporting, where the observed incidence time series are obtained from extremely low to high reporting rates. In Section \ref{Chap4SecBenin}, we illustrate our model by using it to analyze the reported measles incidence in Benin from 2012 to 2018. Section \ref{Chap4SecDisc} contains concluding remarks.


\section{Methods} \label{Chap4SecMethod}

\subsection{Model specification} \label{Chap4SecMod}

In general, disease incidence (i.e., the number of new infections in a time period) is commonly modeled at time steps equal to the average generation time of the disease, which is about 14 days for measles \citep{who_measles2}. To better accommodate the monthly reported incidence data, we model the underlying true measles incidence in semi-monthly intervals. We let $m$ and $t$ index the monthly and the semi-monthly time points, so that $t = 2m-1$ and $t = 2m$ represent the first and second semi-monthly time points in month $m$. In addition, we use $C$ and $I$ to denote the reported and the underlying true measles incidence counts respectively. We account for under-reporting by letting the monthly reported (observed) incidence count $C_m$ follow a Binomial distribution with a sample size equal to the sum of the true semi-monthly incidence in that month, $I_{2m-1} + I_{2m}$, and a constant reporting probability $\rho$, 
\begin{gather}
C_{m} | I_{2m-1}, I_{2m} \sim \mbox{Bin}\left(I_{2m-1}+ I_{2m}, \rho \right). \label{Chap4Eqn1}
\end{gather}

For the underlying (unobserved) true incidence, we let the number of new infections in the current semi-month $I_t$ follow a negative binomial distribution with a mean that depends on the incidence $I_{t-1}$, the susceptibles $S_{t-1}$ and the known total population $N_{t-1}$ from the previous semi-month, 
\begin{gather}
I_{t} | I_{t-1}, S_{t-1} \sim \mbox{NegBin}(\lambda_{t}, \phi) \quad \mbox{for } t \ge 2, \label{Chap4Eqn1.25}\\ \label{Chap4Eqn1.5}
\lambda_{t} = \exp(\beta^{\AR}_t) \frac{I_{t-1}^{\alpha} S_{t-1}}{N_{t-1}} + \exp(\beta^{\EN}) N_{t-1}.
\end{gather}
We assume a constant dispersion parameter $\phi$ and endemic parameter $\beta^{\EN}$ over time. The variance of $I_{t}$ is equal to $\lambda_{t} \left(1 + \frac{\lambda_{t}}{\phi} \right)$, so that smaller value of $\phi$ means more overdispersion. The mixing parameter $\alpha$ is set to be 0.975, which has been justified as accounting for the discrete-time approximation to the continuous time model \citep{glass2003interpreting}. As for the auto-regressive parameter $\beta^{\AR}_t$, we use the following formulation to capture the yearly seasonality of measles transmission, 
\begin{gather}
\beta^{\AR}_t = \gamma_1 + \gamma_2 t + \gamma_3 \mbox{sin} \left(\frac{2 \pi }{\omega} t \right) + \gamma_4 \mbox{cos} \left(\frac{2 \pi}{\omega}  t \right),
\end{gather}
where $\omega = 24$, for semi-monthly incidence time series. 

For the underlying dynamic of the susceptibles, we propose a balancing equation that incorporates the information from past SIA campaigns. Specifically, we model the number of susceptibles in the first semi-month $S_1$ as the product of the proportion of susceptibles in the total population $\theta$ (to be estimated) and the known size of the total population at that time point $N_1$. For semi-month $t \ge 2$, we let the number of susceptibles $S_t$ be equal to the number of susceptibles in the previous semi-month $S_{t-1}$, plus the known number of births entering the susceptible pool $B_{t}$, minus the number of new infections $I_{t}$, and finally minus the number of susceptibles immunized in the most recent SIA campaign $S^*_t$,  
\begin{align} 
S_1 &= \theta \times N_1, \\
S_{t} &= S_{t-1} + B_{t} - I_{t} - S^*_{t}\quad \mbox{for } t \ge 2. \label{Chap4Eqn2}
\end{align} 

The number of births entering the susceptible pool $B_{t}$ can be calculated as the number of 9-month-old children who are losing maternal immunity and are not effectively vaccinated via RI. We call this quantity the \textit{adjusted births}. Let $L_t$ and $R_t$ be the number of 9-month-old children in the population and the RI-specific coverage of the first dose of measles-containing-vaccine (MCV1) among the 9-month-olds at semi-month $t$. The adjusted births $B_{t}$ is obtained as
\begin{gather}
     B_t = L_t \times (1 - R_t \times 0.87), \label{Chap4Eqn3}
\end{gather}
assuming the efficacy of MCV1 in LMIC settings is 87\% \citep{who_measles1}. 

We calculate $S^*_t$, the number of susceptibles immunized in semi-month $t$ in the most recent SIA campaign, as
\begin{gather} 
S^*_{t} = p \times S_{k(t)} \times \delta_t \label{Chap4Eqn3.5}
\end{gather} 
where $p$ denotes the overall SIA efficacy parameter, representing the fraction of susceptible population immunized after a complete SIA campaign, $S_{k(t)}$ denotes the number of susceptibles right before the most recent campaign started, and $\delta_t$ denotes the fraction of the total target population to be covered in semi-month $t$ in the most recent campaign. For any semi-month $t$ in which no SIA activity is carried out, we set $\delta_t = 0$; otherwise, an estimate of $\delta_t$ can be obtained using the target population information from the SIA calendar on the World Health Organization (WHO) website \citep{who_measles3}. This formulation of $S^*_t$ allows for flexible modeling of susceptible dynamics when an SIA campaign is implemented in several phases with part of the total target population covered in each phase (e.g., by geographic regions).


\subsection{Computation} \label{Chap4SecComp}

\citet{morton2005discrete} described a Markov chain Monte Carlo (MCMC) algorithm for making inference about parameters in a TSIR model, but acknowledged that this approach can perform poorly in estimating the reporting rate parameter $\rho$. In preliminary work we also encounter serious unidentifibility issues when we estimate $\rho$ jointly with all other model parameters. Therefore, we propose a two-stage hybrid approach that estimates the reporting rate parameter separately from the rest of the model parameters. Specifically, we estimate the reporting rate parameter in the first stage via an ordinary least squares (OLS) regression model with robust standard errors. The results are then used to estimate the rest of the model parameters in the second stage via MCMC. We now describe each stage in detail. 


\subsubsection{Stage 1: reporting rate estimation via OLS regression with robust standard errors} \label{Chap4SecCompStage1}

The reporting rate estimation step is inspired by the susceptible reconstruction method developed by \citet{finkenstadt2000time}. We consider a time period during which \textbf{no SIA is implemented}. Let the first month of this time period be $m = 1$ and the last month be $m = M$. Let $C_m$, $I_m$, $S_m$ and $B_m$ be the number of reported incidence, underlying true incidence, susceptible population and the adjusted births entering the susceptible pool in month $m$. Given a realization of the underlying monthly true incidence $\{I_m\}$ and the observed monthly reported incidence $\{C_m\}$, for $m = 1, \dots, M$, we let $\rho_m$ be the under-reporting ratio 
\begin{gather*}
\rho_m = \frac{C_m}{I_m} \quad \text{so that} \quad I_m = \frac{1}{\rho_m} C_m = \kappa_m C_m
\end{gather*}
so that $\kappa_m$ is the inverse of $\rho_m$. 

Now, we consider the underlying susceptible dynamics during this time period. Since no SIA is implemented, Equation (\ref{Chap4Eqn2}) implies that the monthly susceptible time series is
\begin{align*}
S_{m} &= S_{m-1} + B_{m} - I_{m} \\
      &= S_{m-1} + B_{m} - \kappa_m C_m  \\
      &= (S_{m-2} + B_{m-1} - \kappa_{m-1} C_{m-1}) + B_{m} - \kappa_m C_m \\ 
      &\hspace{0.25cm}\vdots \\
      &= S_{0} + \sum_{i=1}^m B_{i} - \sum_{i=1}^m \kappa_{i} C_{i} \numberthis{}\label{Chap4Eqn4}
\end{align*}
where $S_0$ is the number of susceptibles right before month 1. Assuming the ratio between the reported incidence and the underlying true incidence $\rho_m$ is roughly constant and equal to $\rho$ during the entire time period, that is, $\rho_m \approx \rho$ and $\kappa_m  = \frac{1}{\rho_m} \approx \frac{1}{\rho} = \kappa$, we can write Equation (\ref{Chap4Eqn4}) as 
\begin{align}
S_{m} &\approx S_{0} + \sum_{i=1}^m B_{i} - \kappa \sum_{i=1}^m C_{i} \label{Chap4Eqn5}
\end{align}
Let $\bar{S}$ be the average of $\{S_m\}$ during the time period $m = 1, \dots, M$. We rewrite Equation (\ref{Chap4Eqn5}) as
\begin{align*}
\sum_{i=1}^m B_{i} &= (S_m - S_0) + \kappa \sum_{i=1}^m C_{i} \\ 
&= (\bar{S} - S_0) + \kappa \sum_{i=1}^m C_{i} + (S_m - \bar{S}) \numberthis{} \label{Chap4Eqn6}
\end{align*}

Let $Y_m = \sum_{i=1}^m B_{i}$ be the cumulative adjusted births, $X_m = \sum_{i=1}^m C_{i}$ be the cumulative reported incidence, $\beta_0 = \bar{S} - S_0$ be the difference between $S_0$ and $\bar{S}$, and $U_m = S_m - \bar{S}$ be the deviation of $S_m$ from the average $\bar{S}$. Then, we can rewrite Equation (\ref{Chap4Eqn6}) as
\begin{gather}
Y_m = \beta_0 + \kappa X_m + U_m \label{Chap4Eqn7}
\end{gather}

Assuming the deviations/errors $U_m$ follow a zero-mean distribution with variance $\sigma^2_m$, Equation (\ref{Chap4Eqn7}) motivates an OLS regression model with robust standard errors \citep{huber1967behavior, white1980heteroskedasticity} for estimating $\kappa$. We can obtain a point estimate $\hat{\kappa}$ and a robust standard error  $\hat{\sigma}_{\kappa}$ for the inverse of the reporting rate $\kappa = \frac{1}{\rho}$ by fitting an OLS regression model with robust standard errors using the cumulative adjusted births $Y_m = \sum_{i=1}^m B_{i}$ as the outcome variable and the cumulative reported incidence $X_m = \sum_{i=1}^m C_{i}$ as the predictor variable. It should be noted that this method shows bias in estimating reporting rate $\rho$ because of the positive dependence between $X_m$ and $U_m$ induced by their relationship through the basic accounting equation. More details regarding this bias can be found in Web Appendix B in the Supporting Information. 


\subsubsection{Stage 2: MCMC for Bayesian hierarchical model} \label{Chap4SecCompStage2}

In the second stage, we estimate the other model parameters and the latent variables $I_t$ using an MCMC algorithm that is similar to the one outlined by \citet{morton2005discrete}. Note that the posterior conditional distribution of any $I_t$ involves all future values of $I$ and the other parameters, because $I_t$ influences all $S_j$, $j > t$, from Equation (\ref{Chap4Eqn2}). To propagate the uncertainty associated with the estimation of $\rho$ from the first stage, we plug in a random value for $\rho$ in each MCMC iteration by taking the inverse of a random sample drawn from the distribution $\mbox{N}\left( \hat{\kappa}, \hat{\sigma}^2_{\kappa} \right)$. 

We use the Metropolis-Hasting algorithm to draw posterior samples and adopt the following guidelines:
\begin{enumerate}
\item We use independent proper diffuse priors, such as the Normal, Uniform or Beta distribution, depending on the range of the parameters. 

\item The starting values for $I_t$ and $\theta$ must be realizable,
i.e., $C_m \le I_{2m-1}+ I_{2m}$ and $I_{t+1} \le S_t \le N_t$ for each $m$ and $t$. Any proposed move such that these rules are violated is rejected. 

\item We update each $I_t$ individually using a discrete proposal distribution as outlined by \citet{wakefield2011bayes}. After updating an $I_t$, all subsequent values in the susceptible series $S_t$ are updated as well. 

\item In preliminary work we saw that the posterior samples of $\theta$, $\gamma_1$ and $\gamma_2$ tend to be highly correlated. We use the blocked Metropolis sampling technique \citep{wakefield2013bayesian} to update them together for more efficient convergence. 
\end{enumerate}


\section{Simulation study} \label{Chap4SecSim}

In this section, we illustrate our method and investigate the effect of uncertainty propagation via a simulation study considering a range of reporting rates, from extremely low to high ($\rho = 0.01, 0.1, 0.3, 0.5 \text{ and }0.7$). 

\subsection{Set up}

We simulated semi-monthly population and live births time series assuming a yearly population growth rate of 2.5\% and a yearly birth rate of 4\%, values that are common among LMICs \citep{UNpop}. To simulate the time series of adjusted births entering the susceptible pool, we assumed an RI-specific MCV1 coverage of 70\% and a first-dose efficacy of 87\% \citep{who_measles1}. We set the starting values of the incidence and susceptibles times series based on a simulation using the Epidemiological MODeling (EMOD) software developed by the Institute of Disease Modeling \citep{idm_emod}, and generated the subsequent $I_t$ and $S_t$ values based on the model described in Section \ref{Chap4SecMethod} using the following true parameter values:
\begin{gather*}
\gamma_1 = 3, \quad \gamma_2 = 0, \quad \gamma_3 = 0.2, \quad \gamma_4 = 0.5, \quad \beta^{\EN} = -12, \quad \phi = 10
\end{gather*}
We allowed the number of infected and susceptible individuals to stochastically change for one year before collecting eight years of semi-monthly time series data. This resulted in 5.6\% of the total population being susceptible to measles at the first time point, i.e., $\theta  = 0.056$. We let there be one SIA with an overall efficacy of $p = 40\%$ implemented in two phases at the beginning of the fourth year, with 50\% of the total target population covered in each phase. We simulated a semi-monthly time series of underlying incidence $I_t$ based on Equations (\ref{Chap4Eqn1.25}) to (\ref{Chap4Eqn3.5}). Then, for each reporting rate, we simulated a monthly observed incidence time series $C_m$ based on Equation (\ref{Chap4Eqn1}). 

We used the simulated data from the first six years as the training set for model fitting and the data from the last two years (i.e., the forecast period) for forecast validation. To evaluate the model's ability to estimate the potential effect of a ``planned" SIA campaign, we simulated an additional set of validation data assuming a second SIA with the same overall efficacy of $p = 40\%$ was implemented at the beginning of the forecast period. 

The simulated total population and adjusted births entering the susceptible population are shown in Web Figure 1 in the Supporting Information. The simulated time series of the underlying incidence and susceptible population, with and without a ``planned" SIA at the beginning of the forecast period, are shown in Web Figures 2 and 3 in the Supporting Information. 


\subsection{Computation and results} \label{Chap4SecSimResult}

\begin{figure}
\centering
\includegraphics[width=0.6\textwidth]{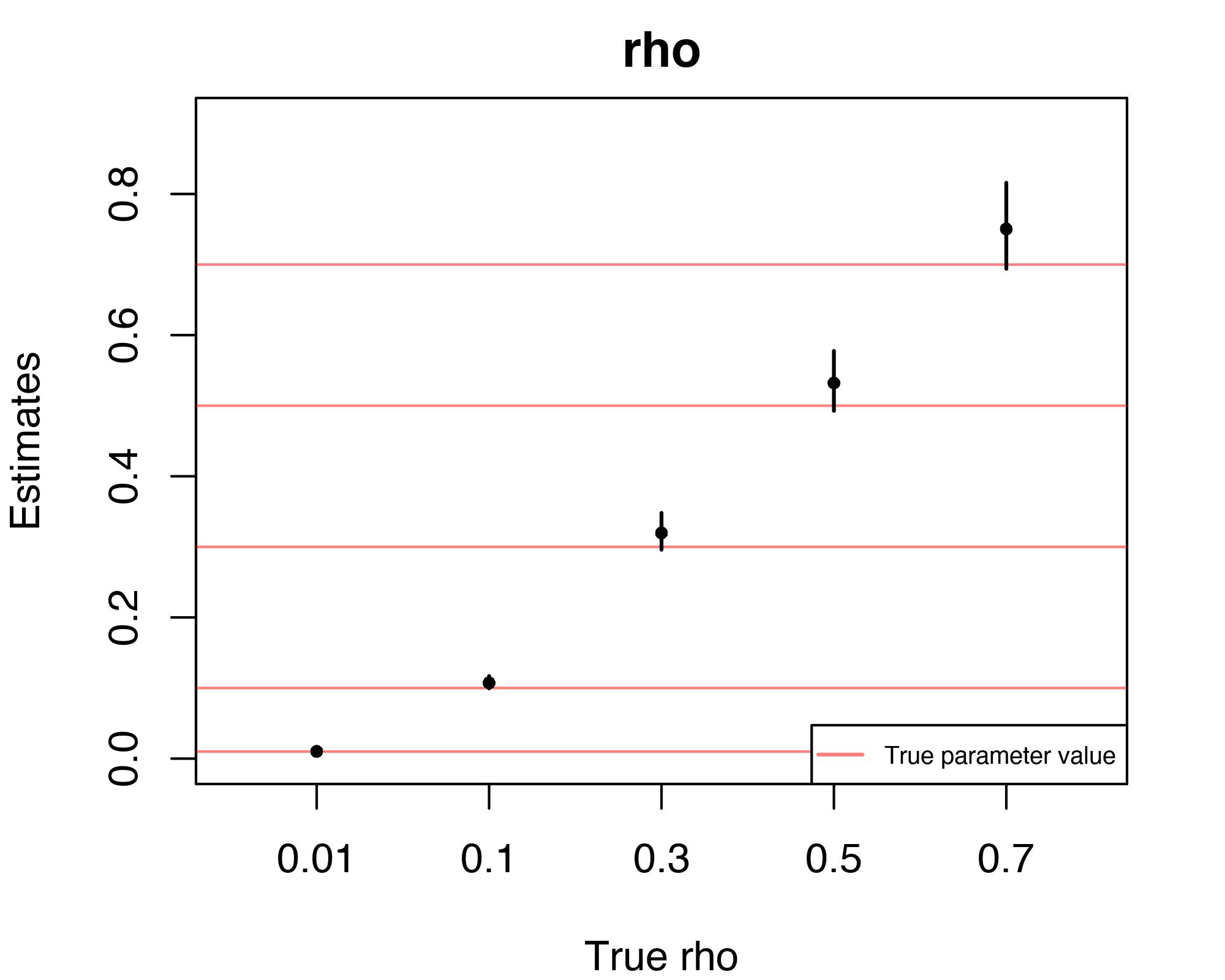}
\caption{The point estimates and 95\% confidence intervals of the reporting rate parameter $\rho$ obtained by fitting OLS regression models with robust standard errors to the cumulative monthly adjusted births and the cumulative monthly reported incidence before the first SIA. The estimates are calculated by drawing 10000 random samples from the distribution $\mbox{N}\left( \hat{\kappa}, \hat{\sigma}^2_{\kappa} \right)$ and taking the inverse of the samples. When $\rho = 0.01$, the point estimate is $0.0102$ and the 95\% confidence interval is $(0.0095, 0.0111)$. When $\rho = 0.1$, the point estimate is $0.107$ and the 95\% confidence interval is $(0.099, 0.117)$.}
\label{Chap4FigSimPar1}
\end{figure}

In each simulation corresponding to a different reporting probability, we estimated the reporting rate $\rho$ in the first stage using the training data from the time period before the first SIA. Figure \ref{Chap4FigSimPar1} shows the point estimates and 95\% confidence intervals of $\rho$ obtained by fitting OLS regression models with robust standard errors to the cumulative monthly adjusted births and the cumulative monthly reported incidence. The estimates are calculated by drawing 10000 random samples from the distribution $\mbox{N}\left( \hat{\kappa}, \hat{\sigma}^2_{\kappa} \right)$ and taking the inverse of the samples. Under all reporting rate scenarios, the point estimate is higher than the true parameter value. There are two issues that we believe are relevant with respect to bias in the reporting rate, one is finite sample bias, and the other is due to the covariance between the error terms and the explanataory variable. The positive bias we are seeing here is likely due to the relatively short length of the monthly time series data used to fit the OLS (36 monthly time points). We conducted an additional simulation study to investigate the behavior of the OLS estimates of the reporting rate using time series of different shapes and lengths (Web Appdendix B) and found that the OLS procedure tends to over-estimate reporting rate $\rho$ when shorter time series are used to fit the regression model. As the length of time series increases, the magnitude of over-estimation generally decreases and reaches 0 at some point. As the length increases further, the procedure produces under-estimated results and the magnitude of under-estimation eventually approaches a constant value, which is due to the dependence between errors and the explanatory variable. More details regarding the bias in the OLS estimation of the reporting rate $\rho$ can be found in Web Appendix B in the Supporting Information.

In the second stage of the estimation, we used all data from the first six years (i.e., the training set) to estimate the other model parameters and the underlying incidence and susceptible dynamics. We drew 20000 posterior samples using the MCMC algorithm and the uncertainty propagation procedure described in Section \ref{Chap4SecCompStage2}, discarded the first 10000 samples as burn-in, and obtained the posterior medians and the 95\% posterior credible intervals (CIs) of parameters. The traceplots of the posterior samples are shown in Web Figures 4 to 8 in the Supporting Information. 

To investigate the impact of uncertainty propagation associated with the reporting rate estimation in the first stage, we carried out an additional round of model fitting using the same training data without the uncertainty propagation step. Specifically, instead of using random samples of $\rho$ drawn from the distribution estimated from the first stage, we plugged in the true reporting rate in each MCMC iteration in the second stage. This would allow us to see how much uncertainty in the model estimation and prediction can be attributed to the uncertainty in reporting rate estimation. The traceplots of the posterior samples obtained from this round of computation are shown in Web Figures 9 to 13 in the Supporting Information. 

\begin{figure}
\centering
\includegraphics[width=0.4\textwidth]{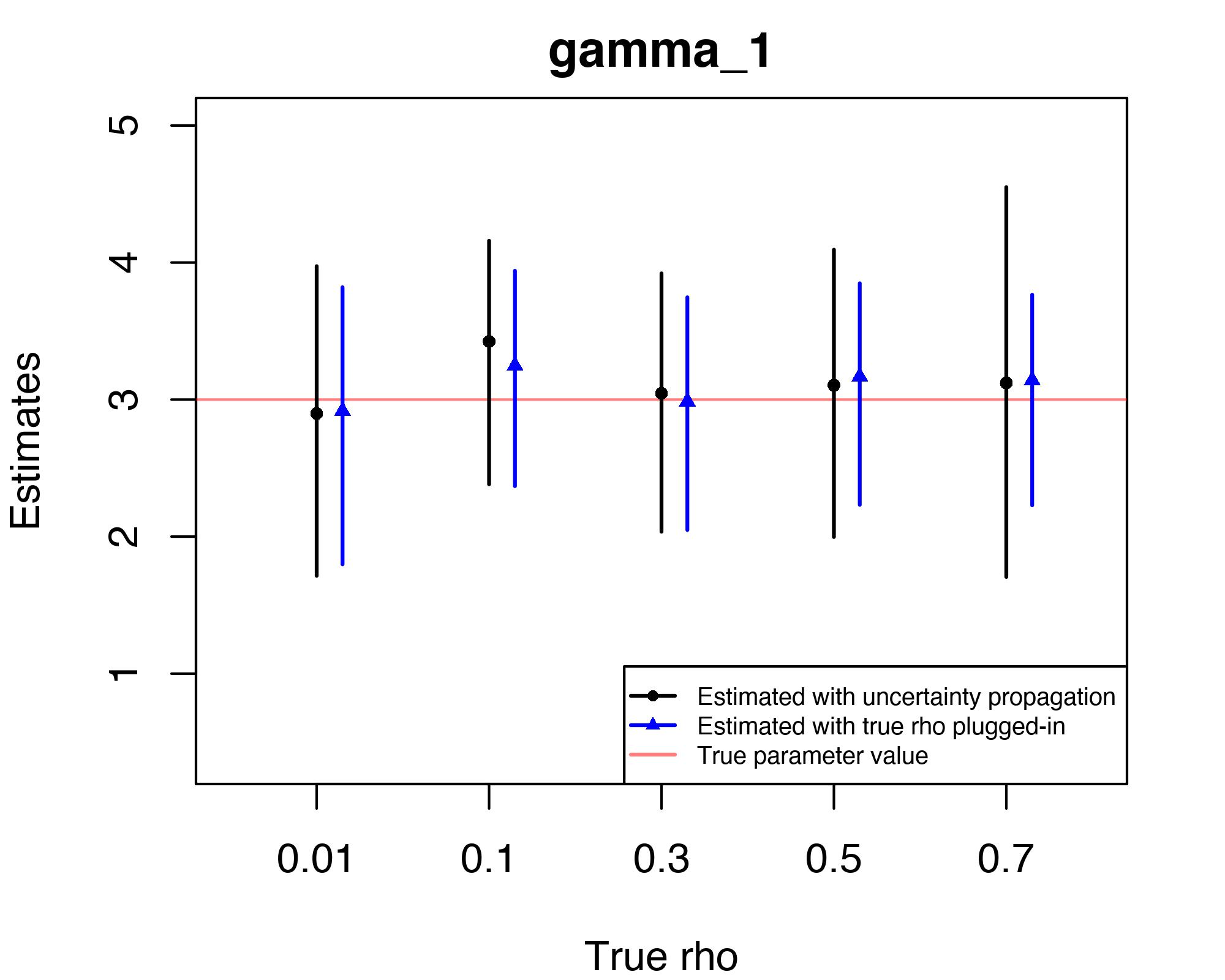}
\includegraphics[width=0.4\textwidth]{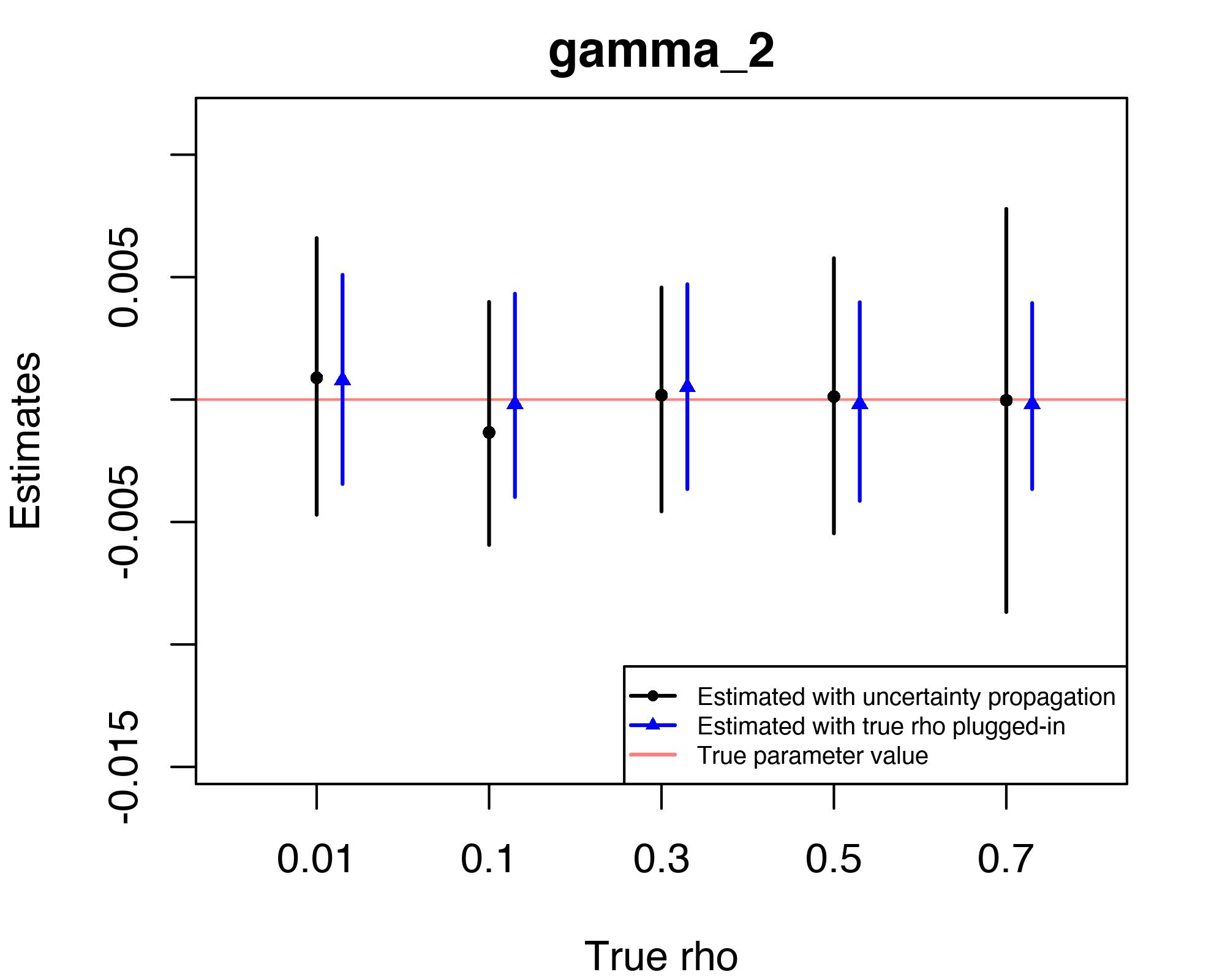}
\includegraphics[width=0.4\textwidth]{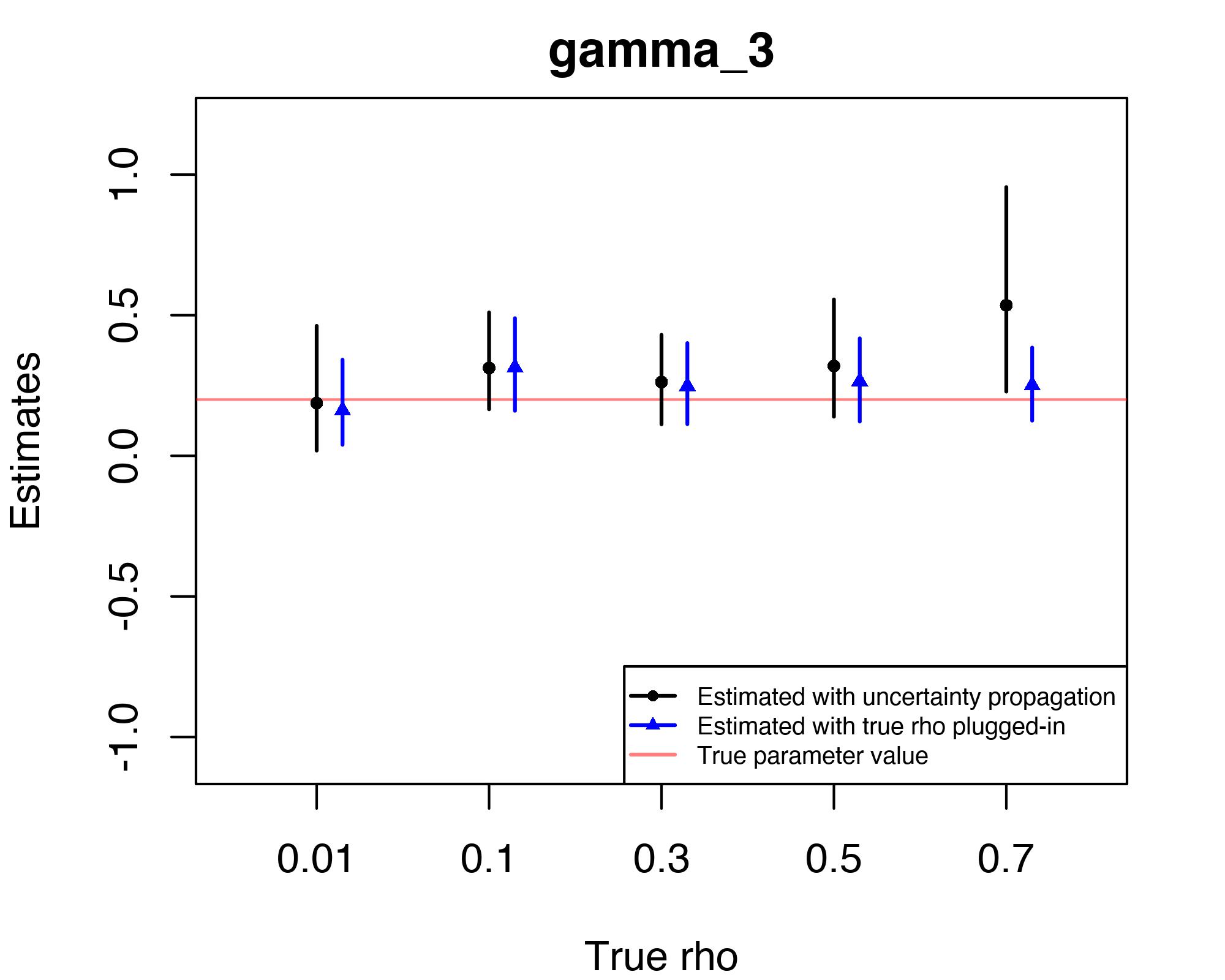}
\includegraphics[width=0.4\textwidth]{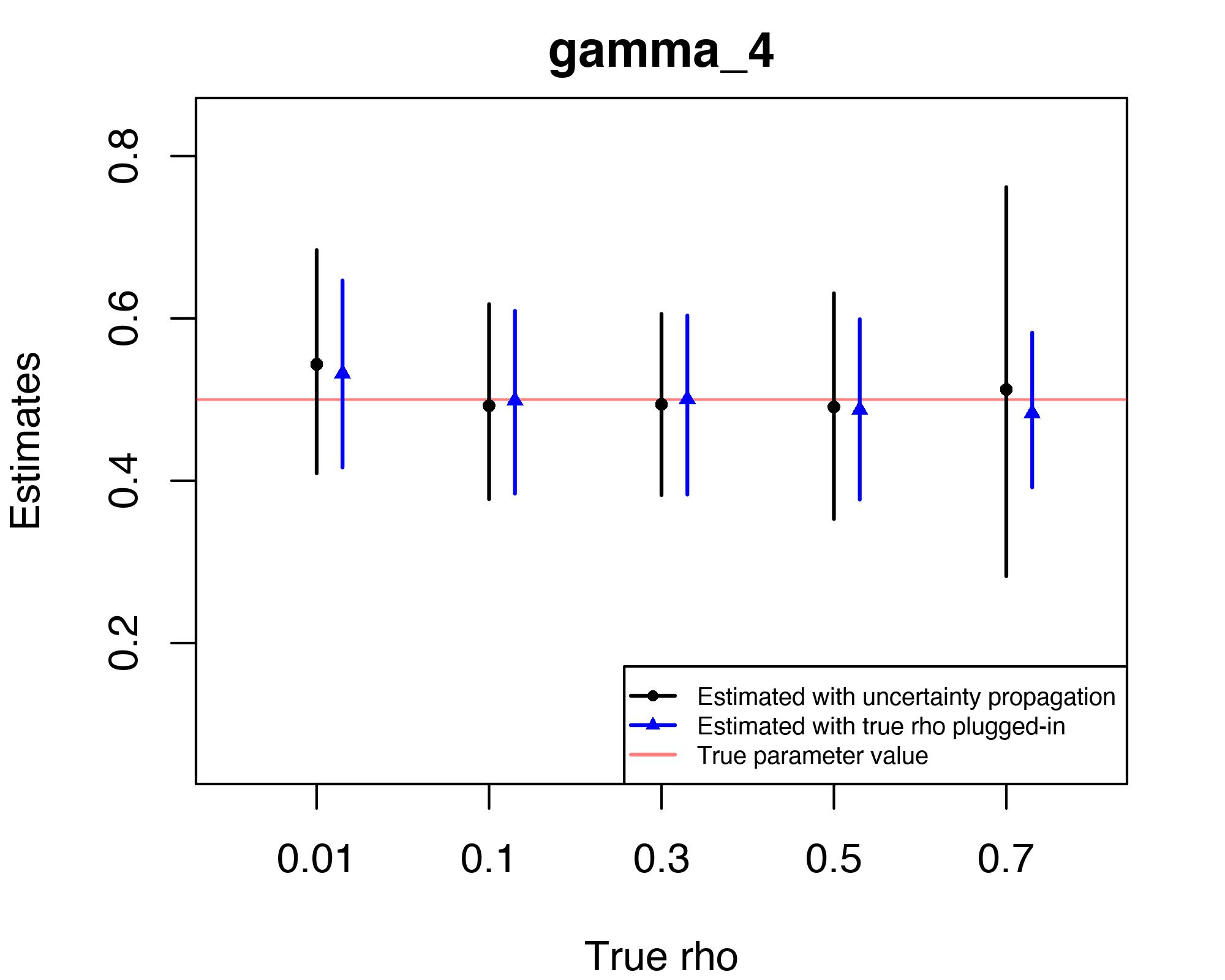}
\includegraphics[width=0.4\textwidth]{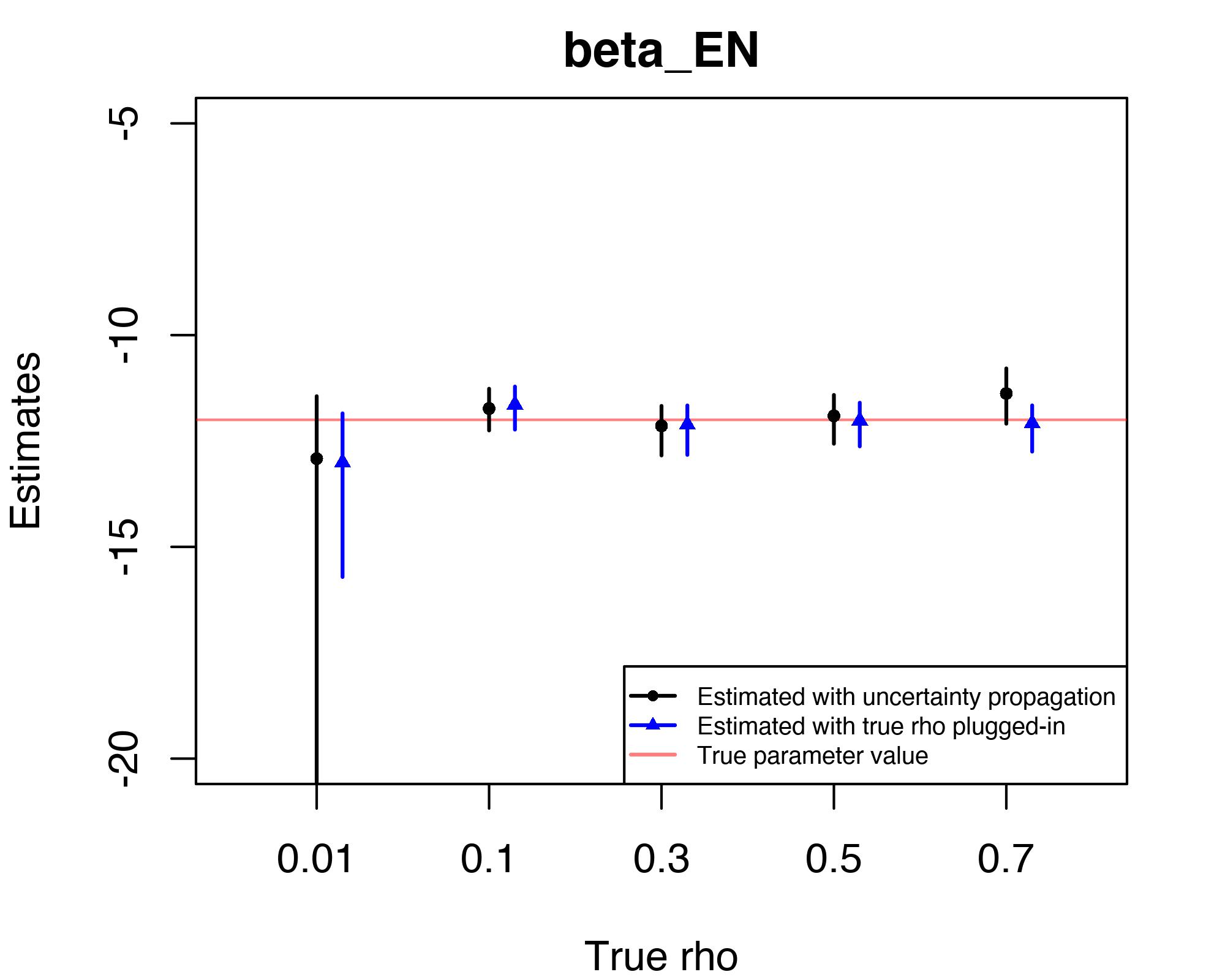}
\includegraphics[width=0.4\textwidth]{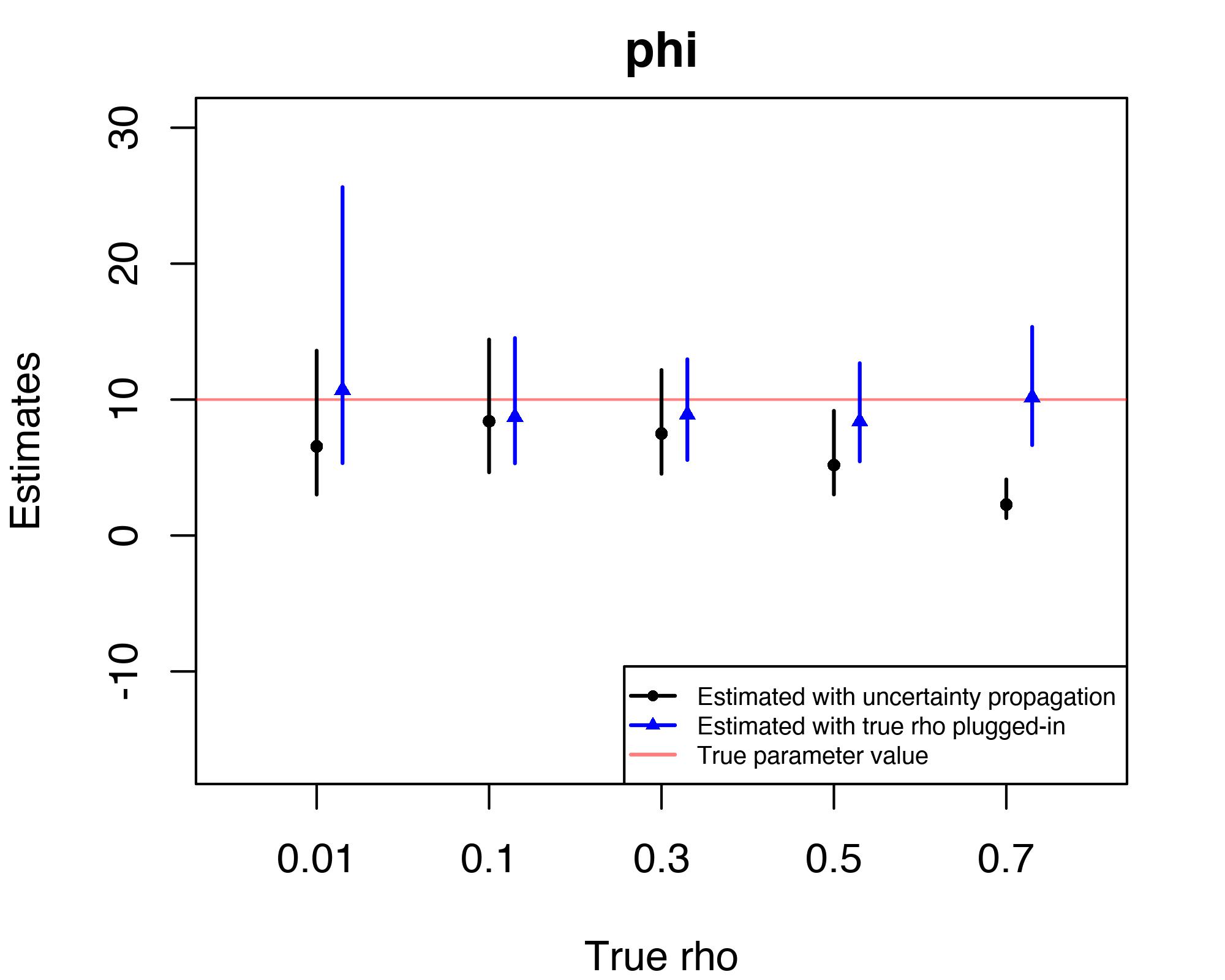}
\includegraphics[width=0.4\textwidth]{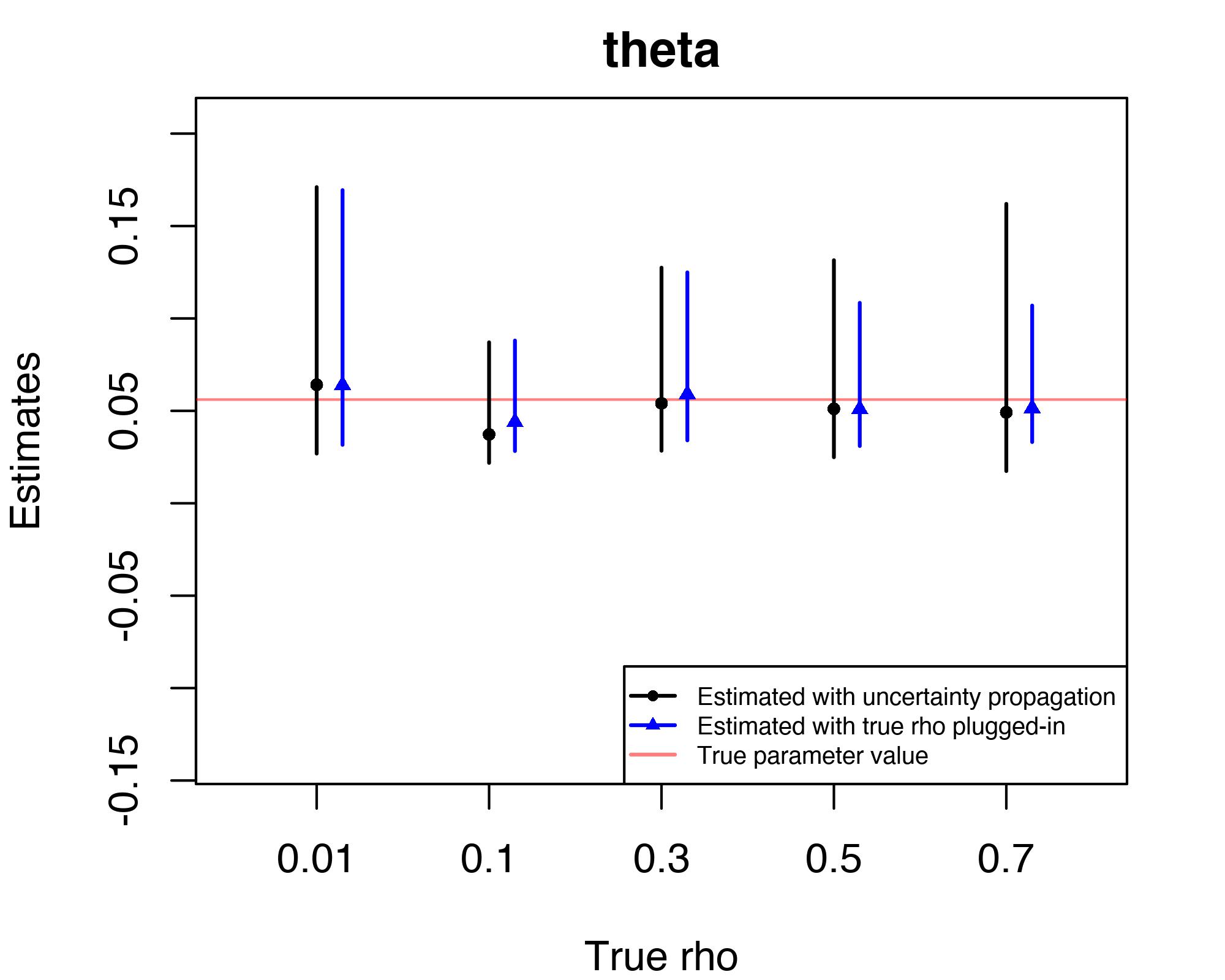}
\includegraphics[width=0.4\textwidth]{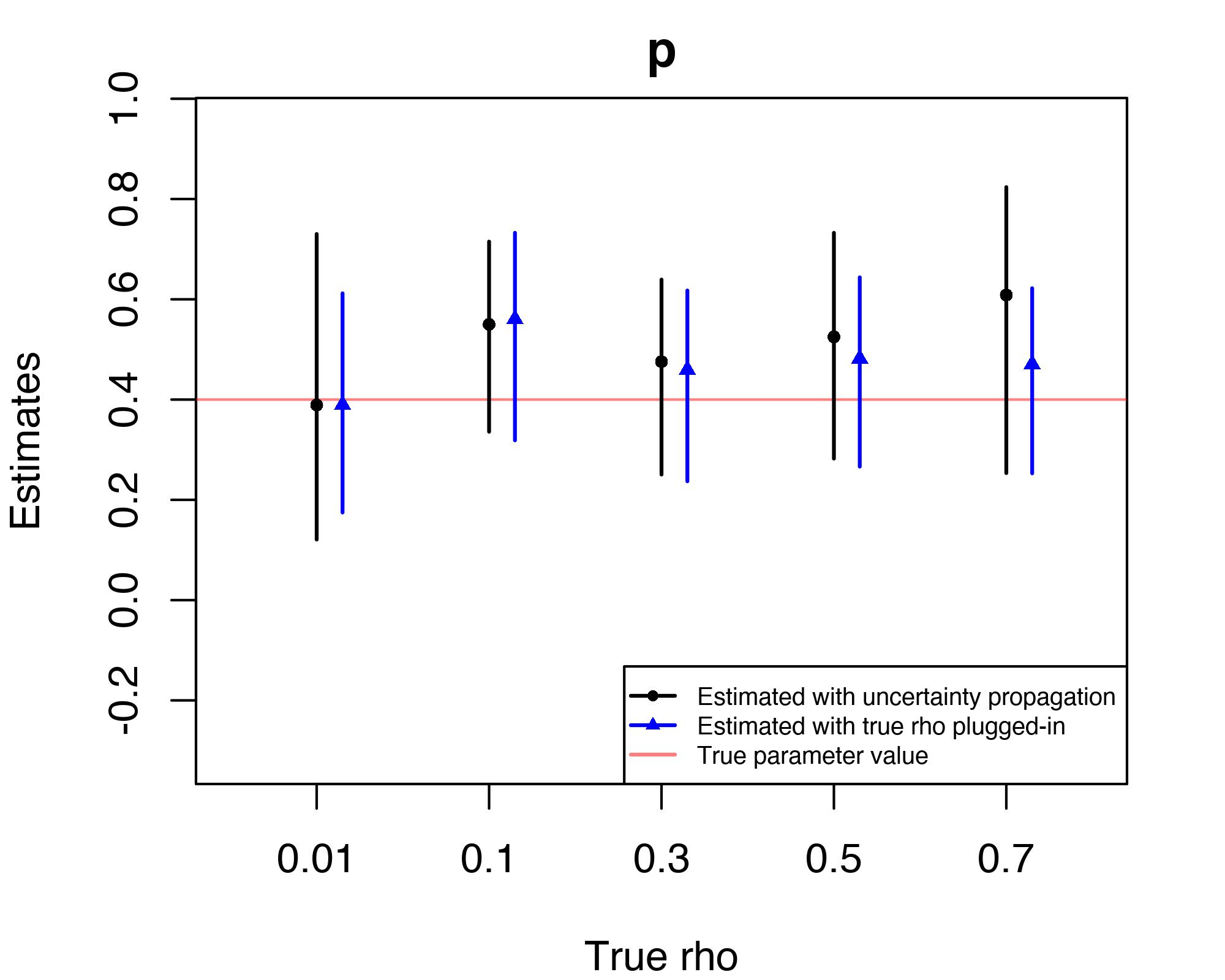}
\caption{The posterior medians and the 95\% posterior credible intervals (CIs) of $\gamma_1$, $\gamma_2$, $\gamma_3$, $\gamma_4$, $\beta^{\EN}$, $\phi$, $\theta$ and $p$ for when the reporting rate is 0.01, 0.1, 0.3, 0.5 and 0.7 (along the x-axis), computed with (black) and without (blue) uncertainty propagation from the reporting rate estimation in the first stage.}
\label{Chap4FigSimPar2}
\end{figure}

Figure \ref{Chap4FigSimPar2} shows the posterior medians and the 95\% posterior CIs of the model parameters computed with and without uncertainty propagation under various reporting rate scenarios. When we estimate the model parameters with the true $\rho$ values plugged in (blue triangles and lines), most of the resultant point estimates are close to the true parameter values and all 95\% posterior CIs cover the truth. The model tends to overestimate the SIA efficacy parameter $p$. Within each parameter, the uncertainties associated with the estimates are quite similar across the various levels of reporting rate, except for when the reporting rate is extremely low ($\rho = 0.01$), in which case the uncertainties tend to be considerably larger than the other scenarios. 

When we estimate the model parameters with the first-stage uncertainty propagated (black triangles and lines), the resultant estimates tend to have wider 95\% posterior CIs than the plugged-in results. The increases in CI widths tend to be larger when the reporting rate is higher. The impact of uncertainty propagation is especially considerable when $\rho = 0.7$, in which case the uncertainty associated with the first-stage estimation of $\rho$ is the greatest compared to the other scenarios. One exemption we see is the estimation of the dispersion parameter $\phi$, in which case the uncertainty propagation resulted in considerable decreases in both the point estimates and the posterior CI widths. One possible reason is that some of the variability in the underlying incidence induced by the uncertainty of the $\rho$ estimates was attributed to overdispersion, which lead to lower estimated values of the dispersion parameter $\phi$ (corresponding to greater overdisperson). 

To predict the underlying incidence and susceptible population dynamics in the forecast period, we used the posterior samples of the model parameters from each MCMC iteration and generated the potential realizations of the underlying dynamics according to Equations (\ref{Chap4Eqn1.5}) through (\ref{Chap4Eqn3.5}). For the simulations where a second SIA was implemented at the beginning of the forecast period, we carried out the prediction assuming the ``planned" SIA has the same (estimated) efficacy of the previous campaign. 

Web Figures 14 and 15 in the Supporting Information show the simulated true values, posterior medians and 95\% CIs/predictive intervals, and 200 randomly selected posterior samples of the underlying measles incidence under various reporting rate scenarios, computed with and without uncertainty propagation and/or a ``planned" SIA at the beginning of the forecast period. The corresponding results for the underlying susceptible population dynamics are shown in Web Figures 16 and 17 in the Supporting Information. Under all reporting rate scenarios, the model successfully predicted the general timing and the relative magnitudes of the two measles outbreaks in the forecast period when no SIA is implemented, and captured the effect of a ``planned" SIA if it were implemented at the beginning of the forecast period. However, the model was not able to predict the peaks of the two outbreaks accurately. The results shown in Web Figures 14 and 15 in the Supporting Information also demonstrate the impact of uncertainty propagation in incidence forecasting. Under all reporting rate scenarios, we see that the predictive intervals became wider after the first-stage estimation uncertainty is accounted for in the second stage, and the magnitude of the change increases with the first-stage uncertainty. 

Finally, we computed the root mean squared error (RMSE) of the predicted underlying incidence during the forecast period, $\mbox{RMSE}_f$, to assess the model's forecast accuracy: 
\begin{gather*}
    \mbox{RMSE}_f = \sqrt{ \frac{1}{T_f} \sum_{t=1}^{T_f}\left(\hat{I}_t - I_t \right)^2},
\end{gather*}
where $I_t$ and $\hat{I}_t$ are the ``true" simulated incidence and the posterior median of the predicted incidence at time $t$, and  $T_f = 48$ is the total number of semi-monthly time steps in the forecast period. 

\begin{figure}
\centering
\includegraphics[width=0.6\textwidth]{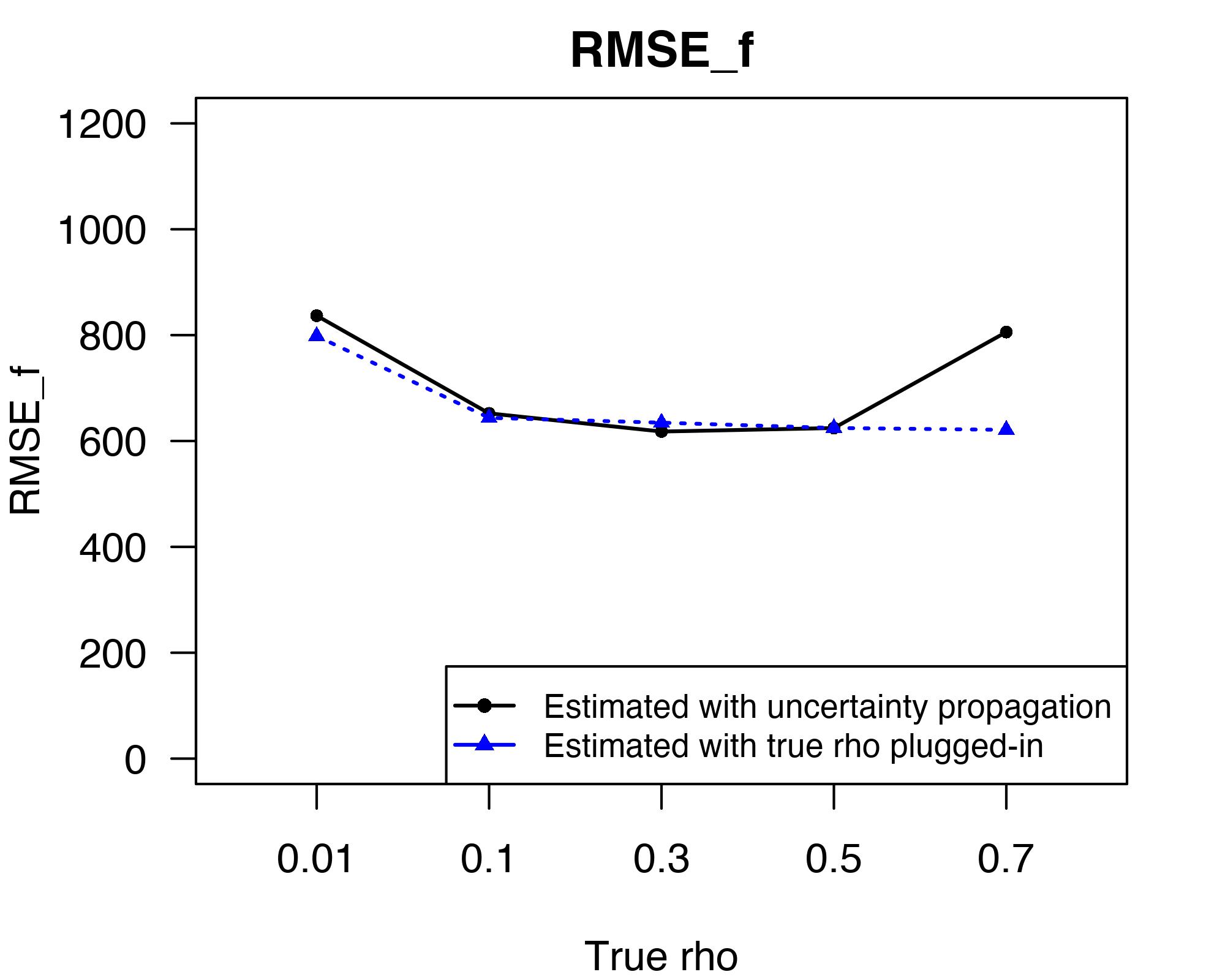}
\caption{The root mean squared error of the underlying incidence in the forecast period, $\mbox{RMSE}_f$, calculated using the posterior medians of the semi-monthly incidence estimated with (black) and without (blue) uncertainty propagation from the reporting rate estimation in the first stage. }
\label{Chap4FigSimRMSE}
\end{figure}

Figure \ref{Chap4FigSimRMSE} shows the results computed with (black) and without (blue) uncertainty propagation from the reporting rate estimation in the first stage. When the true $\rho$ is plugged-in in the second stage estimation, the $\mbox{RMSE}_f$ is the highest when the reporting rate is extremely low ($\rho = 0.01$). As reporting rate increased, the forecast accuracy measured by the $\mbox{RMSE}_f$ first improved sharply, and then remained relatively stable. The impact of uncertainty propagation on $\mbox{RMSE}_f$ is small under under reporting rate scenarios in general, except for when the reporting rate is high ($\rho = 0.7$) and the fist-stage uncertainty and bias are large. In this case, the $\mbox{RMSE}_f$ value computed with uncertainty propagation is much higher than the plugged-in value. 

\section{Analysis of the measles incidence data in Benin} \label{Chap4SecBenin}

In this section, we illustrate our method by analyzing the monthly reported measles incidence in Benin between 2012 and 2018. Specifically, we fit our model using data collected between January 2012 and December 2016, estimate the efficacy of the one SIA campaign carried out in this time period, and forecast the reported incidence between January 2017 and December 2018 for model validation. 

\subsection{Data}

We obtained the yearly national population estimates and crude birth rates of Benin from 2011 to 2018 from the WorldBank database \citep{WorldBank}. The estimates were linearly interpolated over time to calculate the semi-monthly total population and live-births time series on the national level. To obtain the adjusted births entering the susceptible pool, we first calculated the department-level (administrative level 1 in Benin) live-births using a set of proportions derived from the WorldPop estimates \citep{WorldPopBenin}. We used the infant mortality rate estimates from the Institute for Health Metrics and Evaluation (IHME) \citep{IHMEBenin} to approximate the under-9-month mortality rate in each department and calculated the number of 9-month-old children at each time point. Using the space-time smoothing model developed by \citet{dong2020space}, we also obtained the department-level RI-specific MCV1 coverage with data from the Multiple Indicator Cluster Surveys (MICS) conducted in 2006, 2012 and 2017 \citep{MICS} and the Demographic and Health Survey (DHS) conducted in 2014 \citep{DHS}. Finally, we calculated the adjusted births time series for each of the 12 departments in Benin assuming the efficacy of the first dose of measles vaccine is 87\% \citep{who_measles1} based on Equation (\ref{Chap4Eqn3}), and aggregated them to obtain the adjusted births on the national level. 

The monthly reported measles incidence and the SIA calendar were downloaded from the WHO database \citep{who_measles3}. There was one national SIA campaign implemented in two phases during the time period of interest: 2.6 million children aged between 9 months to 9 years were first targeted at the beginning of December 2014, then another 0.4 million children of the same age group were targeted at the end of January 2015. The reported incidence time series is shown in Web Figure 18 in the Supporting Information. 


\subsection{Computation and results}

For reporting rate estimation, we fitted an OLS regression model with robust standard errors using the cumulative adjusted births and cumulative reported incidence data collected between January 2012 and December 2014. We drew 10000 random samples from the distribution $\mbox{N}\left( \hat{\kappa}, \hat{\sigma}^2_{\kappa} \right)$ and took the inverse of the samples to obtain estimates for the reporting rate parameter $\rho$. The resultant estimates for $\rho$ are extremely low, with a point estimate of 0.0074 and a 95\% confidence interval of (0.0069, 0.0081). This suggested that there was severe under-reporting in Benin's measles surveillance data during this time period. 

We carried out the second stage of estimation described in Section \ref{Chap4SecCompStage2} with uncertainty propagation using data collected between January 2012 and December 2016. We drew a total of 20000 posterior samples, discarded the first 10000 samples as burn-in, and obtained the posterior medians and the 95\% posterior CIs of the model parameters and the underlying incidence time series. The traceplots of the posterior samples are shown in Web Figure 12 in the Supporting Information.  

Table \ref{Chap4TabBeninSumm} shows the summary results for the parameter estimation. In particular, the estimates for $\theta$ implied that an estimated 2.8\% (95\% CI: 0.5\%, 14.9\%) of the total population was susceptible to measles at the beginning of year 2012. As for SIA efficacy, the point estimate of $p$ suggested that an estimated 49.9\% of the susceptible population were immunized after the SIA implemented in 2014--15. However, the 95\% posterior CI of $p$ is very wide (14.5\%, 85.9\%), indicating that the estimated efficacy was associated with huge uncertainty. This is expected considering the extremely low reporting rate estimates from the first stage and the relatively short time series we used for this analysis. 

\begin{figure}
\centering
\includegraphics[width=0.7\textwidth]{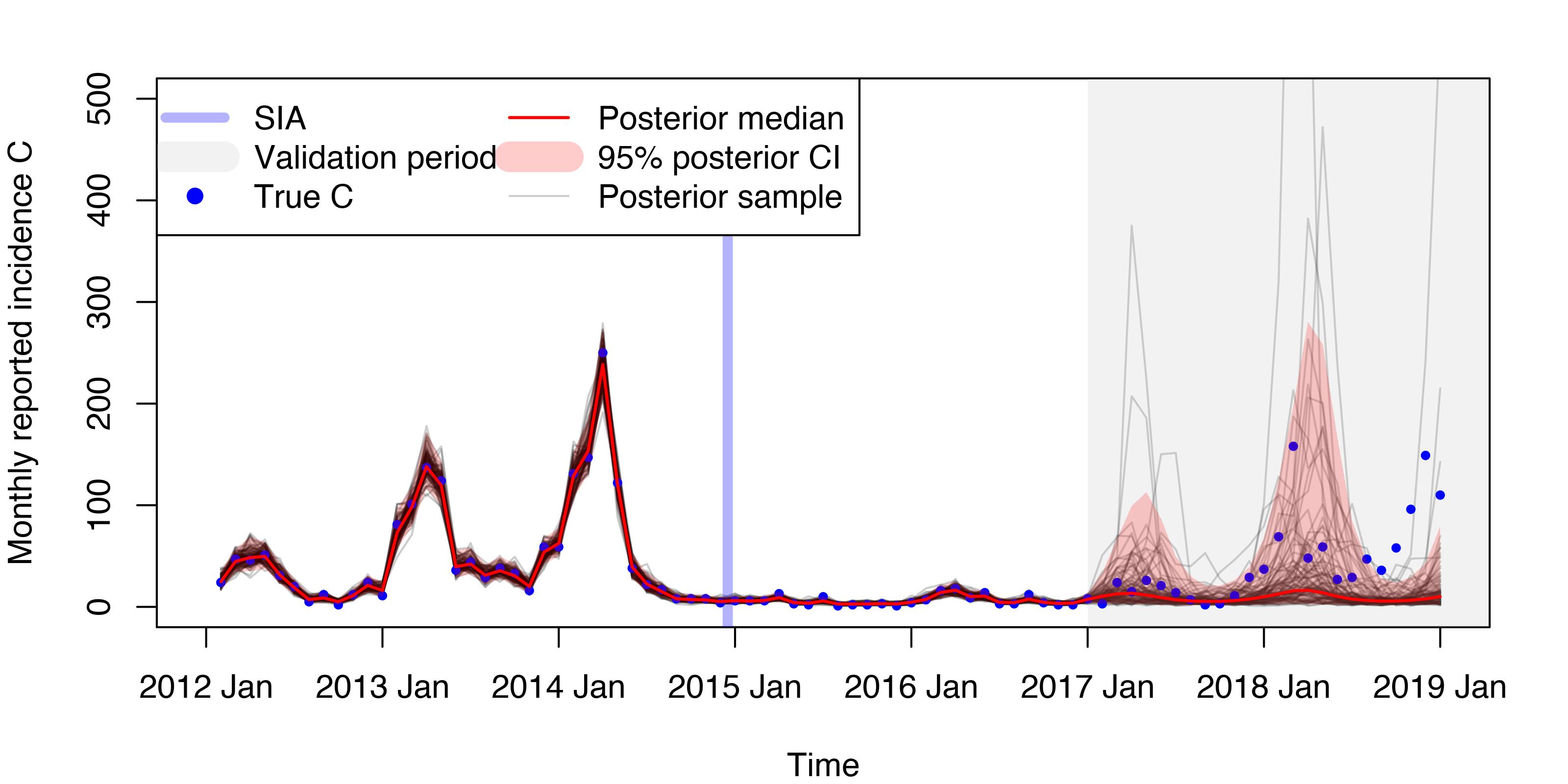}\\
\includegraphics[width=0.7\textwidth]{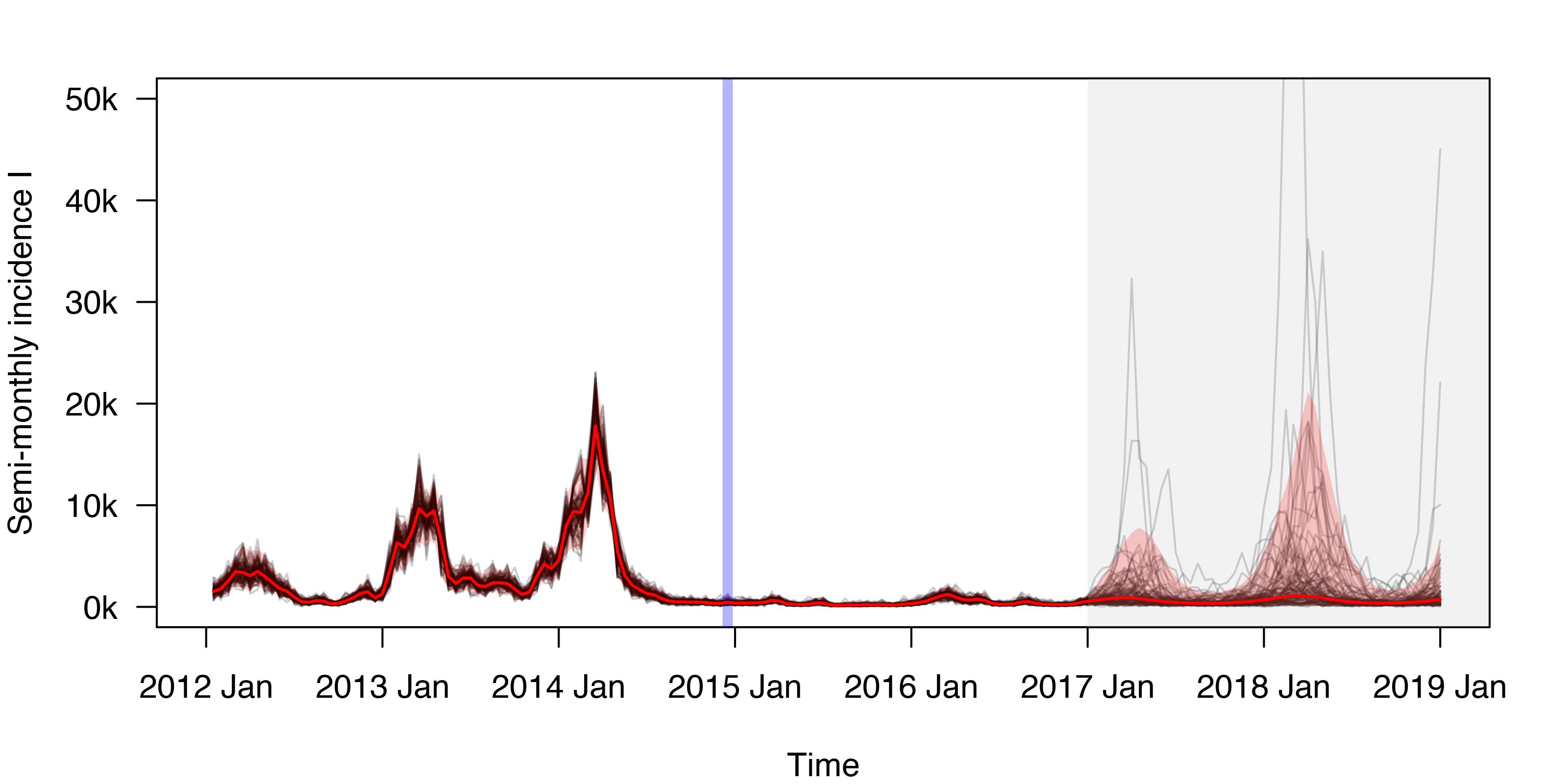} \\
\includegraphics[width=0.7\textwidth]{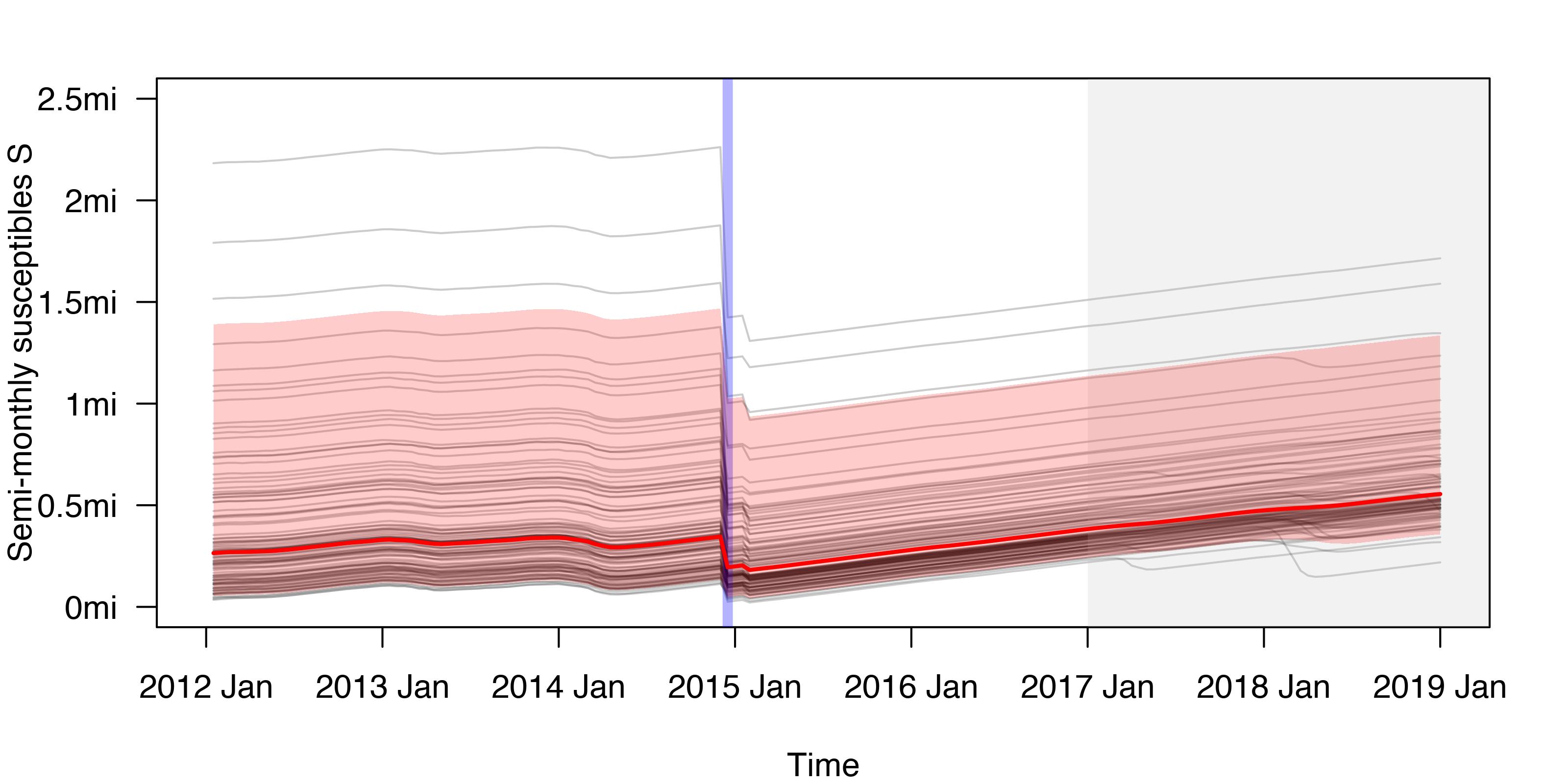} \\
\caption{The posterior medians, the 95\% posterior CIs/predictive intervals and 200 randomly selected posterior samples of (1) Top: the underlying monthly measles incidence scaled by the estimated reporting rate (with the true reported incidence time series); (2) Middle: the underlying semi-monthly true incidence; and (3) Bottom: the underlying semi-monthly susceptible population in Benin between 2012 and 2018.}
\label{Chap4FigBeninPred}
\end{figure}

Finally, we predicted the underlying incidence and susceptible population dynamics in the forecast period by generating the potential realizations of the underlying dynamics using the posterior samples of the model parameters from all MCMC iterations. In Figure \ref{Chap4FigBeninPred}, we compare the posterior medians, 95\% CIs/predictive intervals, and 200 randomly selected posterior samples of the underlying monthly measles incidence, scaled by the estimated reporting rate, to the reported incidence time series in Benin between 2012 and 2018. The model was able to predict the timing of the first two outbreaks and their relative magnitudes in the forecast period, but failed to predict the precise timing or magnitude of the third outbreak. The estimated underlying incidence and susceptible population dynamics are also shown in Figure \ref{Chap4FigBeninPred} for reference. 

\begin{table}
\centering
\caption{The posterior medians and the 95\% posterior CIs of model parameters from the Benin analysis. The estimates for $\rho$ were obtained by fitting an OLS regression model with robust standard errors to the cumulative monthly adjusted births and the cumulative monthly reported incidence before December 2014. The estimates are calculated by drawing 10000 random samples from the distribution $\mbox{N}\left( \hat{\kappa}, \hat{\sigma}^2_{\kappa} \right)$ and taking the inverse of the samples. }
\label{Chap4TabBeninSumm}
\begin{tabular}{ccc}
  \hline \hline
Parameter & Estimate & 95\% CI  \\ 
  \hline
  $\rho$ & 0.0074 & (0.0069, 0.0081) \\
  $\gamma_1$ & 3.08 & (2.10, 5.34) \\ 
  $\gamma_2$ & -0.005 & (-0.016, 0.003) \\
  $\gamma_3$ & 0.173 & (-0.002, 0.376) \\
  $\gamma_4$ & 0.248 & (0.080, 0.438) \\
  $\beta^{\EN}$ & -11.2 & (-12.5, -10.5) \\ 
  $\theta$ & 0.028 & (0.005, 0.149) \\ 
  $p$ & 0.499 & (0.145,	0.859) \\ 
  $\phi$ & 4.89 & (2.48, 9.87) \\ 
   \hline \hline
\end{tabular}
\end{table}

\section{Discussion} \label{Chap4SecDisc}

In this chapter, we developed a discrete-time hidden Markov model under the TSIR framework to estimate SIA efficacy using measles incidence time series data. Our approach accounts for under-reporting and seasonality of measles transmission, and accommodates monthly reported incidence data that is publicly available from the WHO database. We proposed a two-stage estimation procedure that first estimates the reporting rate through an OLS regression model with robust standard errors and then estimates the rest of the model parameters and the underlying incidence dynamics via an MCMC algorithm with uncertainty propagation. The proposed method can be used to estimate the fraction of susceptible people immunized by past SIAs and forecast incidence trends in the future under various hypothetical SIA scenarios. 

We illustrated our method and investigated the impact of uncertainty propagation on parameter estimation and incidence prediction via a simulation study considering a range of reporting rates. In addition, we also applied our method to analyze the reported measles incidence data in Benin from 2012 to 2018. While both the simulation study and the Benin example demonstrate our model's ability to predict the general timing and relative magnitudes of measles outbreaks in the near future, they also revealed some of the limitations of our approach and the TSIR framework. 

As mentioned in Sections \ref{Chap4SecCompStage1} and \ref{Chap4SecSimResult}, the OLS procedure produces biased results in reporting rate estimation. We investigated the sources and properties of the bias in an additional simulation study and presented the results in Web Appendix B of the Supporting Information. We found that the OLS procedure generally over-estimates the reporting rate parameter $\rho$ when short time series are used to fit the regression model. This is likely a result of the small sample size of the data. As the length of time series increases, the magnitude of over-estimation generally decreases and reaches 0 when certain length is reached. However, as the length increases further, the procedure starts to under-estimate $\rho$ and the magnitude of under-estimation eventually approaches a constant value. This negative bias is likely a result of the positive correlation between the cumulative reported incidence and the deviance of the susceptible time series from its mean. Our investigation is the first to uncover and derive the sources of the bias in this popular regression-based method for reporting rate estimation \citep{ferrari2008dynamics, metcalf2011epidemiology, metcalf2013implications, mahmud2017comparative,  thakkar2019decreasing}. We are now actively working on a method for correcting this bias.

Another aspect of our model that could be improved is to allow temporal fluctuations in the reporting rate that may be influenced by time-varying factors such as the state of the epidemic and reports in the media. \citet{finkenstadt2000time} described a spline-based method for estimating time-varying reporting rates, but the approach uses an ad-hoc algorithm for deciding the bandwidth and gave poor performance in some of our preliminary work. As \citet{noufaily2019under} has pointed out, under-reporting corrections should be geographical- and disease-specific, as well as age- and gender-dependent. We are currently exploring alternative model specifications that can potentially incorporate additional information, such as measures of availability and reliability of surveillance system, to detect and account for temporal trends in reporting rates. 

In our model, the semi-monthly time series of total population and adjusted births are input data that are treated as the truth --- we consider them to be deterministic quantities with fixed values. A potential extension of our current method is to account for potential uncertainties in the population and births data and propagate these uncertainties in parameter estimation and incidence prediction. In addition, the balancing equation characterizing the susceptible dynamic currently assumes negligible non-measles deaths among the susceptible people. To relax this assumption, one can potentially add another component to the balancing equation to explicitly account for the people exiting the susceptible pool due to non-measles deaths. The viability and data requirement of this extension still requires further investigation. 

It should be noted that our model was derived based on some key underlying assumptions that the TSIR models require. These include homogeneous mixing of the human host population and \textit{frequency-dependent} contact rate (see Web Appendix A in the Supporting Information for details). Therefore, the model is not suitable for analyzing data from extremely large geographical areas in which there is considerable heterogeneity in measles transmission dynamics. An interesting future research topic is to extend the current method to jointly model incidence from multiple areas while allowing for spatially and temporally correlated transmission patterns.

\backmatter

\section*{Acknowledgements}

Jon Wakefield was supported by grant R01 AI029168 from the National Institutes of Health, Tracy Qi Dong by grant
U54 GM111274 from the National Institutes of Health. The authors are grateful to Dr. Kevin McCarthy, Dr. Niket Thakkar, and Dr. Kurt Frey from Institute for Disease Modeling for their informative discussions and data preparations. 


\section*{Supporting Information}

Additional supporting information may be found online in the Supporting Information section at the end of the article. 


\section*{Data Availability Statement}

The data used in this paper to support our findings are available from the corresponding author upon reasonable request.


\bibliographystyle{biom}
\bibliography{References}


\label{lastpage}
\end{document}